\documentclass[twocolumn,showpacs,amsmath,amssymb,aps,pra,floatfix,nofootinbib,superscriptaddress]{revtex4}
\usepackage{graphicx,graphics,times,color}
\usepackage{bm}
\usepackage{longtable}
\usepackage{color}
\usepackage{lipsum}

\def\be{\begin{equation}}
\def\ee{\end{equation}}
\def\ba{\begin{eqnarray}}
\def\ea{\end{eqnarray}}
\def\la{\langle}
\def\ra{\rangle}

\def\h{\hskip 1cm}

\begin{document}


\title{Measurement induced dynamics for spin chain quantum communication and its application for optical lattices}

\author{Sima Pouyandeh}
\affiliation{Department of Physics, Isfahan University of Technology, Isfahan 84156-83111, Iran}

\author{Farhad Shahbazi}
\affiliation{Department of Physics, Isfahan University of Technology, Isfahan 84156-83111, Iran}

\author{Abolfazl Bayat}
\affiliation{Department of Physics and Astronomy, University College London, Gower St., London WC1E 6BT, United Kingdom}

\date{\today}

\begin{abstract}
We present a protocol for quantum state transfer and remote state preparation across spin chains which operate in their anti-ferromagnetic mode. The proposed mechanism harnesses the inherent entanglement of the ground state of the strongly correlated many-body systems which naturally exists for free. The uniform Hamiltonian of the system does not need any engineering and, during the whole process, remains intact while a single qubit measurement followed by a single-qubit rotation are employed for both encoding and inducing dynamics in the system. This, in fact, has been inspired by recent progress in observing spin waves in optical lattice experiments, in which manipulation of the Hamiltonian is hard and instead local rotations and measurements have become viable. The attainable average fidelity stays above the classical threshold for chains up to length $~ 50$ and the system shows very good robustness against various sources of imperfection.
\end{abstract}


\maketitle

\section{Introduction}

Strongly correlated many-body systems often have highly entangled nontrivial ground states. The dynamics of such systems can be used for propagating information \cite{bose03} across distant sites and has been studied intensively in the last decade \cite{bose-review,bayat-review-book}. Very recently, experimental realization of quantum state transfer through the natural dynamics of many-body systems have been achieved in NMR \cite{state-transfer-NMR} and coupled optical fibers in linear optics \cite{kwek-perfect-transfer}. Most of the proposals so far (see \cite{bose-review,bayat-review-book} and the references therein), with very few exceptions like \cite{bayat-densecoding}, are based on attaching an extra qubit, which encodes an ``unknown" quantum state, to a chain of strongly interacting particles  which is usually initialized to its ground state unless for certain engineered XX chains in which local end-chain operations makes it to work for any initialization \cite{DiFranco-initialization}. This mode of transmission does not seek to harness the intrinsic entanglement of many-body systems and the symmetries of the Hamiltonian seems to be more important \cite{bayat-xxz}. Moreover, attaching and detaching a single qubit to a many-body system is practically hard and needs a very fine control over the interaction of particles which is missing in many physical systems such as cold atoms.  Although at the receiver site taking the quantum state for further process ultimately may need a swap operator or equivalently controlling some local interactions, for encoding the quantum state at the sender site not demanding such fine control will simplify the fabrication significantly.

One can also think of sending a ``known" quantum state from the sender to receiver. This occurs in a few occasions such as the remote quantum state preparation \cite{Bennet-remote} in which preparing the quantum state at some place is impossible due to practical issues, like inaccessibility of certain sites. Thus, the quantum state has to be prepared at one location and then transferred to the less accessible ones. There might be also several users for whom the quantum states are prepared in a single location (which needs simpler fabrication) and then distributed between them (see Refs.~\cite{bayat-router,Kay-perfect} for more details on information routers). In all these cases \emph{known} quantum states have to be transferred from one place to another.  It may be argued that by knowing the quantum state, the sender can simply send the Bloch vector $(n_x,n_y,n_z)$ of the qubit to the receiver via classical communication to prepare the state at the receiver site and there is no need for quantum communication. This possibility is indeed correct, however, the parameters of the Bloch vector are real numbers and sending them may need very long string of classical bits which may not be desired and has to be compensated by loosing precision in using a shorter set of classical bits. A single quantum state, however, can take all that information in a single shot. Hence, sending known quantum states, either considered as state transfer or remote state preparation, has its own merit while has hardly been studied for spin spin chain communication \cite{bayat-densecoding}.

Quantum measurement is one of the mysteries of physics which has been hardly understood since the birth of quantum mechanics. According to quantum theory, measuring any observable results in a \emph{random} output which is one of the eigenvalues of a Hermitian operator that is associated to that particular observable. The probability of such an outcome is determined by the overlap of the initial wave function and the corresponding eigenvector of the observable operator. In fact, after the measurement the wave function of the system goes under an abrupt change and collapses to that particular eigenstate of the observable operator. So far, the quantum measurement has been exploited for quantum communication via teleportation \cite{Bennet-teleportation} and measurement-based quantum computation \cite{Briegel-MBQC}.  In conventional spin chain quantum communications, however, the random nature of measurement has been an obstacle for incorporating it in quantum state transfer protocols.  On the other hand, since in quantum measurement the state of the system collapses instantaneously it can be used to induce dynamics in the system by changing its state and thus may be used as an alternative approach to attaching scenarios for quantum communication in strongly correlated systems.

Cold atoms in an optical lattice are excellent test bed for many-body experiments. Both bosons \cite{Mott-insulator-boson} and fermions \cite{Mott-insulator-fermion} have been realized in the Mott insulator phase, where there is exactly one atom per site, and by properly controlling the intensity of laser beams one can tune the interaction between neutral atoms to behave as an effective spin Hamiltonian \cite{Lukin}. Local addressability of atoms with the resolution of single sites \cite{Bloch-single-site1,Bloch-single-site2} has opened a totally new window for exploring many-body systems. Single site unitary operations and measurements \cite{Bloch-single-site2,Meschede,Bloch-spin-wave,Bloch-magnon} are in fact becoming viable and accessible with high fidelities. Thanks to these new advancements, the correlated particle-hole pairs and string orders \cite{Bloch-particle-hole} together with their time evolution \cite{Bloch-correlation-time} have been explored experimentally. Furthermore, in recent experiments the propagation of a single impurity spin \cite{Bloch-spin-wave} and magnon bound states \cite{Bloch-magnon} in a ferromagnetic spin chain have been investigated.
New cooling techniques  \cite{Medley2010} have enabled, reaching for the first time, the temperatures required for observing quantum magnetic phases emerged due to spin interactions. In view of these, it is very timely to put forward new proposals which are doable with current achievements in cold atom experiments. In particular, one may think of new ways for quantum communication across a strongly correlated many-body interacting systems.

In this paper, we introduce a mechanism for exploiting the inherent entanglement of many-body systems for quantum communication across a spin chain. The encoding of information is done through a single qubit measurement followed by the operation of a unitary gate which is determined by the random outcome of the measurement. The following measurement induced dynamics propagates the quantum state through the chain till it reaches the other side in which the information is captured by switching off the interaction couplings.  The proposed protocol, which has been inspired by recent achievements for observing spin waves in ferromagnetic chains in optical lattices \cite{Bloch-spin-wave,Bloch-magnon}, exploits quantum measurement in order to induce quench dynamics in the system and can be seen as the first step for observing spin dynamics in anti-ferromagnetic chains. The simplicity of the protocol, with all its ingredients available in optical lattice experiments, allows for the experimentation of the proof of principles for measurement induced dynamics along an anti-ferromagnetic chain. Our protocol can also be interpreted as remote state preparation \cite{Bennet-remote} since a known quantum state is prepared on one side of the chain and then is transferred to the other side which might be inaccessible for some practical issues. In addition, our measurement induced transport can serve as information router in which the quantum state is prepared at one site of a network and then distributed between multiple users to reduce the complexity of fabrication.

The structure of the paper is as following. In section \ref{sec2} the model is introduced, in section \ref{sec3} the unrestricted measurement induced dynamics is introduced, in section \ref{sec4} the proposal for restricted measurement is discussed and in section \ref{sec5} entanglement distribution is analyzed. Then in section \ref{sec6} odd chains which do not have SU(2) symmetry are investigated and imperfections are studied in section \ref{sec7}. In section \ref{sec8} the application of our mechanism in optical lattices is explored. Finally in section \ref{sec9} we summarize our results.

\begin{figure} \centering
    \includegraphics[width=8cm,height=6cm,angle=0]{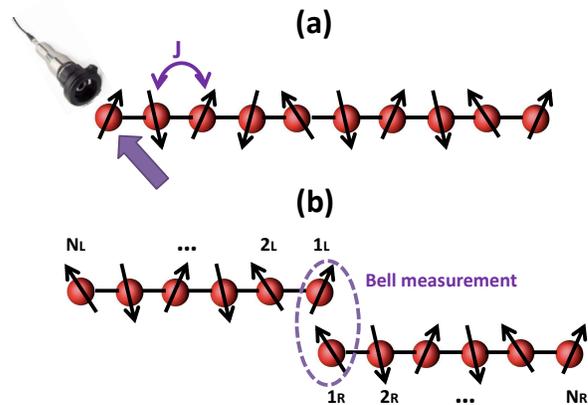}
    \caption{ (Color online) (a) An arrays of interacting qubits for which the interaction type is Heisenberg  with the exchange coupling $J$. A local control is available for the first qubit to operate a quantum gate or perform a spin measurement. (b) A Bell measurement on the first qubits of two noninteracting chains (note that labeling of the atoms are reversed in each chain) is used for entanglement distribution along the two spin chains.  }
     \label{fig1}
\end{figure}

\section{Introducing the model} \label{sec2}

We assume a system of $N$ spin-1/2 particles interacting via an anti-ferromagnetic Heisenberg Hamiltonian
\begin{equation}\label{H}
    H=\sum_{k=1}^{N-1} J_k \overrightarrow{\sigma}_k.\overrightarrow{\sigma}_{k+1}
\end{equation}
where $\overrightarrow{\sigma_k}=(\sigma_k^x,\sigma_k^y,\sigma_k^z)$ is the vector of Pauli operators acting on site $k$ and $J_k$ is the exchange coupling which is assumed to be uniform (i.e. $J_k=J>0$ for all $k$'s) unless it is stated. A schematic picture of this system is shown in Fig. \ref{fig1}(a). System is cooled down to its ground state $|GS\ra$. For the moment we consider even chains (i.e. even $N$) in which due to the SU(2) symmetry of the Hamiltonian the ground state is unique and lies in the subspace that half of the spins are up. Moreover, in even chains the SU(2) symmetry implies that the reduced density matrix of each spin is maximally mixed. This allows us to write the the ground state $|GS\ra$ in a very generic form of
\begin{equation}\label{gs}
    |GS\ra=\frac{|\uparrow_k,\Downarrow\ra-|\downarrow_k,\Uparrow\ra}{\sqrt{2}},
\end{equation}
where $\uparrow_k$ ($\downarrow_k$) means site $k$ is in spin up (spin down) and $\Uparrow$ represents a quantum state for the rest of the system in which there are $N/2$ spins up and $N/2-1$ spins down (similarly for $\Downarrow$ there are $N/2$ spins down and $N/2-1$ spins up). The detailed structure of $|\Uparrow\ra$ and $|\Downarrow\ra$ are very complex and due to their different parities these two states are orthogonal.  In addition due to the the SU(2) symmetry of the system the generic form of the ground state in Eq.~(\ref{gs}) remains valid for any basis of spins.

By measuring a single spin at site $k$, in an arbitrary basis, the quantum state of the whole system collapses according to the outcome of the measurement. For instance, if the measurement is in the $\sigma^z$ basis on site $k$ then with probability of 1/2 the outcome of measurement is spin up and the quantum state of the system collapses to $|\uparrow_k,\Downarrow\ra$. This new state still remains in the subspace of the ground state but is no longer an eigenvector of the Hamiltonian and as the results system evolves under the action of the Hamiltonian $H$. However, the outcome of the measurement is a random process and cannot be used directly for quantum communication across the spin chain. In the rest of the paper we try to exploit the random measurement induced dynamics for the purpose of quantum communication.

\section{Quantum State Transfer: Unrestricted Basis} \label{sec3}

In this section we assume that a general projecting measurement, in any arbitrary basis, is possible at the sender site, which is taken to be site 1. This measurement followed by a conditional unitary operation, which depends on the outcome of the measurement, are used to initialize our desired quantum state in the sender spin. The following unitary time evolution of the system transfers this quantum state to the receiver site.

The most general pure quantum state can be written as
\begin{equation}\label{psi_p1}
    |\psi^{(+1)}\ra=\cos{(\frac{\theta}{2})}|\uparrow\ra+e^{i\phi}\sin{(\frac{\theta}{2})}|\downarrow\ra
\end{equation}
where $0\leq \theta \leq \pi$ and $0\leq \phi \leq 2 \pi$ are the two angles in the spherical coordinates which determine a single point on the surface of the Bloch sphere. This state is the eigenvector of the Hermitian operator $\overrightarrow{\sigma}.\overrightarrow{{\bf n}}$ (with eigenvalue $+1$) where the unit vector $\overrightarrow{{\bf n}}$ is defined as $\overrightarrow{{\bf n}}=(\sin(\theta)\cos(\phi),\sin(\theta)\sin(\phi),cos(\theta))$. The other eigenvector corresponding to the negative eigenvalue (with eigenvalue $-1$) is
\begin{equation}\label{psi_m1}
    |\psi^{(-1)}\ra=\cos{(\frac{\theta}{2})}|\downarrow\ra- e^{-i\phi} \sin{(\frac{\theta}{2})}|\uparrow\ra.
\end{equation}
one can transfer one of these eigenvectors to another by using the following unitary operator
\begin{equation}\label{R_unitary}
R_u=|\psi^{(+1)}\ra \la \psi^{(-1)}|+|\psi^{(-1)}\ra \la \psi^{(+1)}|.
\end{equation}

To initialize the quantum state $|\psi^{(+1)}\ra$ in the sender site we measure the Hermitian operator $\overrightarrow{\sigma}.\overrightarrow{{\bf n}}$ at site 1. With probability of 1/2 the outcome is $+1$ and the initialization is done otherwise with probability of 1/2 the output is $-1$ and thus the unitary operator $R$ should act on site 1 to convert its state into $|\psi^{(+1)}\ra$. As the result of this measurement the quantum state of the whole system changes accordingly. Depending on the outcome of the measurement the quantum state of the system initialized to one of the following states
\begin{eqnarray} \label{initi1_0}
  |\Psi^+(0)\ra &=& \sqrt{2} P^{(+1)} |GS\ra \cr
  |\Psi^-(0)\ra &=& \sqrt{2} R_u P^{(-1)} |GS\ra
\end{eqnarray}
where $P^{(\pm 1)}=|\psi^{(\pm 1)}\ra \la\psi^{(\pm 1)}|$ are the projecting operators and $\sqrt{2}$ is the normalization factor. Each of these states are obtained by probability of 1/2 and as it is clear the unitary operation $R_u$ acts only when the outcome of the measurement is $|\psi^{(-1)}\ra$.

Since neither of these states are the eigenvector of the Hamiltonian they evolve as
\begin{equation}\label{psi_t}
    |\Psi^\pm (t)\ra=e^{-iHt} |\Psi^\pm(0)\ra.
\end{equation}
By tracing out all spins except the receiver, which is taken to be the last spin $N$, one can get the density matrix of received state
\begin{equation}\label{rho_N_t}
    \rho_N^{\pm}(t)=Tr_{\widehat{N}} |\Psi^\pm (t)\ra \la \Psi^\pm (t)|.
\end{equation}
To quantify the quality of state transfer one can compute the fidelity as
\begin{equation}\label{Fid1_t}
    F_u^{\pm}(t)=\la \psi^{(+ 1)}|  \rho_N^{\pm}(t) |\psi^{(+ 1)}\ra.
\end{equation}
Thanks to the SU(2) symmetry of the system $F_u^{\pm}(t)$ is independent of $\theta$ and $\phi$ which means that all quantum states are transferred by the same fidelity.  A general proof for this statement is given in Appendix A.

\begin{figure} \centering
    \includegraphics[width=9cm,height=7cm,angle=0]{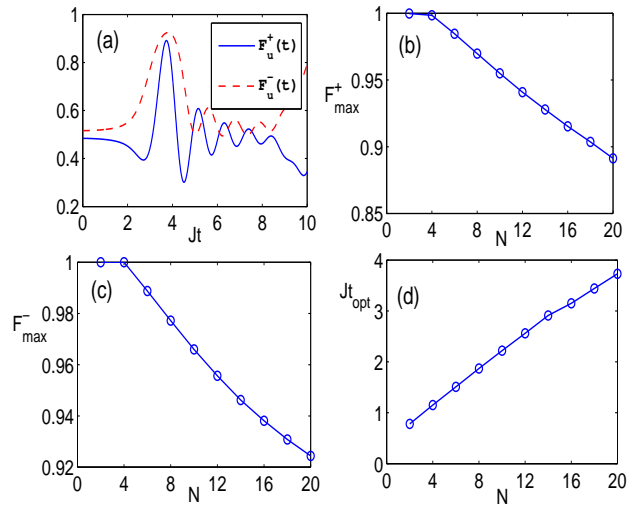}
    \caption{(Color online) (a)  The two fidelities $F_u^{+}(t)$ and $F_u^{-}(t)$ in terms of $Jt$ in a chain of length $N=20$ for the unrestricted measurement protocol. (b) The maximal fidelity $F_{max}^+$ as a function of $N$. (c) The maximal fidelity $F_{max}^-$ as a function of $N$. (d) The optimal time $Jt_{opt}$  versus length $N$.}
     \label{fig2}
\end{figure}

In Fig.~\ref{fig2}(a) the fidelity $F_u^{+}(t)$ and $F_u^{-}(t)$ are both plotted as functions of time. As it is clear from the figures the fidelity starts evolving after a certain time that information reaches the last site. Then due to constructive quantum interferences at a particular time $t=t_{opt}$ the information reaches the receiver site and fidelity peaks for the first time. Though the later peaks might be larger it is physically unwise to wait for such long times as in practical cases the interaction with environment and its induced decoherence deteriorates the quality of transmission. So that we focus on the first peak at which the fidelity takes its maximal value, i.e. $F_{max}^{\pm}=F_u^{\pm}(t_{opt})$.

In Figs.~\ref{fig2}(b) and (c) the maximal fidelities $F_{max}^{+}$ and $F_{max}^{-}$ are plotted versus length $N$. As it is clear from these figures the fidelities are both high and go down almost linearly with very small slopes. A linear fit to data shows that $F_{max}^{+}=-0.007N+1.024$ and $F_{max}^{-}=-0.005N+1.016$. One can use these linear fits to extrapolate the fidelities in longer chains which shows that for chains up to $N \sim 50$ the fidelities are still above the classical threshold 2/3. This indeed shows the very high potential of this strategy for quantum state transfer across a many-body system. In Fig.~\ref{fig2}(d) the optimal time $t_{opt}$ is plotted versus $N$ which also shows a linear dependence on $N$.

\section{Quantum State Transfer: Restricted Basis} \label{sec4}

Very often due to practical issues it is not possible to accomplish quantum measurement in any arbitrary basis on a single spin as needed in the encoding of the previous section. Instead quantum projecting measurement may be possible only for a certain basis, let say $\sigma_z$. The outcome of the measurement is thus either $|\uparrow\ra$ or $|\downarrow\ra$ and the quantum state of the whole system collapses to $|\uparrow\Downarrow\ra$ or $|\downarrow\Uparrow\ra$ respectively. To initialize the spin into a general superposition like Eq.~(\ref{psi_p1}) a further unitary operation on first site is needed. Depending on the outcome of the measurement we apply one of the following unitary operators to the first spin
\begin{eqnarray} \label{R1_R2}
  R_\uparrow &=& |\psi^{(+1)}\ra \la \uparrow | + |\psi^{(-1)}\ra \la \downarrow | \cr
  R_\downarrow &=& |\psi^{(-1)}\ra \la \uparrow | + |\psi^{(+1)}\ra \la \downarrow |
\end{eqnarray}
where $R_\uparrow$ ($R_\downarrow$) is applied if the outcome of the measurement in the $\sigma_z$ basis is $|\uparrow\ra$ ($|\downarrow\ra$) to rotate it to $|\psi^{(+1)}\ra$. The resulted states are not eigenstates of the Hamiltonian $H$ and thus system evolves accordingly. At any time $t$ one can see that the quantum state of the system is one of the following states depending on the measurement result
\begin{eqnarray} \label{initi2_0}
  |\Psi^\uparrow(t)\ra &=&  e^{-iHt}R_\uparrow \otimes I |\uparrow\Downarrow\ra, \cr
  |\Psi^\downarrow(t)\ra &=&  e^{-iHt}R_\downarrow \otimes I |\downarrow\Uparrow\ra,
\end{eqnarray}
As before we compute the density matrix of the last spin by tracing out the rest
\begin{equation} \label{rho_n_ud_t}
 \rho_N^{\alpha}(t)=Tr_{\widehat{N}} |\Psi^\alpha (t)\ra \la \Psi^\alpha (t)| \ \ \ \text{for $\alpha=\uparrow,\downarrow$}.
\end{equation}
To quantify the quality of the state transfer we compute the fidelity as
\begin{equation} \label{fid_r_ud_t}
 F_r^\alpha(t)= \la \psi^{(+1)}|  \rho_N^{\alpha}(t) |\psi^{(+1)}\ra.
\end{equation}
Unlike the fidelity $F_u^\pm (t)$ for unrestricted measurement basis the $F_r^\alpha(t)$ depends on input parameters $\theta$ and $\phi$. To have an input independent quantity one may compute the average fidelity for all possible pure input states on the surface of the Bloch sphere
\begin{equation} \label{Fav_ud_t}
 F_{av}(t)= \frac{1}{4\pi} \int{F_r^\alpha(t) \sin(\theta) d\theta d\phi}.
\end{equation}
Using a little bit of maths one can show that
\begin{widetext}
\begin{eqnarray} \label{Fav_t}
 F_{av}(t)&=& \frac{1}{6} \left\{ \la\downarrow\Downarrow|e^{+iHt}  |\uparrow_N\ra \la\uparrow_N|  e^{-iHt} |\downarrow\Downarrow\ra
           +                  \la\uparrow\Downarrow|e^{+iHt} |\downarrow_N\ra \la\downarrow_N|  e^{-iHt} |\uparrow\Downarrow\ra  \right\} \cr
          &+& \frac{1}{3} \left\{  \la\uparrow\Downarrow|e^{+iHt} |\uparrow_N\ra \la\uparrow_N| e^{-iHt} |\uparrow\Downarrow\ra
          +                 \la\downarrow\Downarrow|e^{+iHt} |\downarrow_N\ra \la\downarrow_N| e^{-iHt} |\downarrow\Downarrow\ra \right\}  \cr
          &+& \frac{1}{3} abs\left\{ \la\downarrow\Downarrow|e^{+iHt} |\downarrow_N\ra \la\uparrow_N| e^{-iHt} |\uparrow\Downarrow \ra \right\}.
\end{eqnarray}
\end{widetext}
where in the above formula it is assumed that the outcome of the measurement is spin up and to have the formula for the outcome spin down one has to only replace $\Downarrow$ with $\Uparrow$ in Eq.~(\ref{Fav_t}). In fact, due to the symmetries of the system $F_{av}(t)$ is identical for both $\alpha=\uparrow,\downarrow$ and thus we drop the index $\alpha$.

\begin{figure} \centering
    \includegraphics[width=9cm,height=5cm,angle=0]{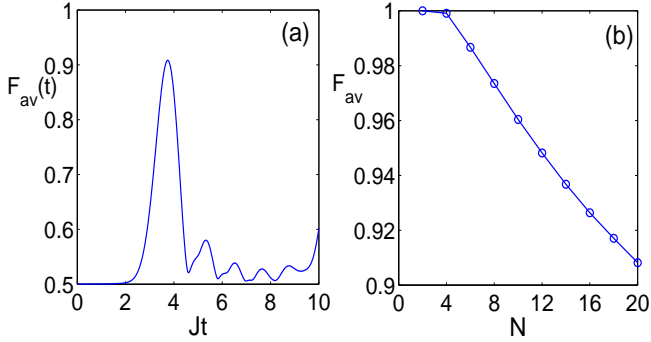}
    \caption{(Color online) (a) The average fidelity $F_{av}(t)$ as a function of $Jt$ for a chain of length $N=20$ in a restricted basis protocol. (b) The maximal average fidelity $F_{av}(t_{opt})$ in terms of length $N$.}
     \label{fig3}
\end{figure}

In Fig.~\ref{fig3}(a) we plot $F_{av}(t)$ as a function of time. At $t=t_{opt}$ the average fidelity peaks for the first time. In Fig.~\ref{fig3}(b) the maximum of average fidelity is depicted in terms of $N$ which can be well fitted by a linear function as $F_{av}(t_{opt})= -0.006 N+1.020$. This shows that for chains up to length $N\approx 60$ the average fidelity is above the classical threshold  2/3.

\begin{table*} \label{table_I}
\begin{centering}
\begin{tabular}{|c|c|c|c|c|c|c|c|c|c|}
  \hline
  $ N $                & 4    & 6    & 8    & 10    & 12    & 14    & 16    & 18    & 20 \\
  \hline
  $F_{av}(projection)$ & 0.9991 & 0.9867 & 0.9735 & 0.9604 & 0.9482 & 0.9368 & 0.9264 & 0.9171 & 0.9082 \\
  \hline
  $F_{av}(attaching)$  & 0.9554 & 0.9212 & 0.8986 & 0.8826 & 0.8693 & 0.8584 & 0.8496 & 0.8425 & 0.8365 \\
  \hline
\end{tabular}
\caption{A comparison between the attainable average fidelity from our proposed projection mechanism (in the restricted basis) and the widely studied attaching scenarios for different lengths.  }
\par\end{centering}
\end{table*}

For the sake of completeness we compare the attainable average fidelity of our proposed mechanism for the restricted basis with the widely studied attaching procedures, in which one extra qubit that carries our desired quantum state is attached to a spin chain initialized in its ground state just as Ref.~\cite{bayat-xxz}. The results have been given in TABLE I and as it is clear from the data the projective mechanism gives higher fidelity in comparison to the attaching scenarios. The same sort of improvement is observed for the unrestricted projective measurement (not shown in the TABLE I).

\section{Entanglement Distribution} \label{sec5}

The proposed measurement induced dynamics for state transfer can also be used for entanglement distribution. To fulfill such task we consider two independent chains which do not interact with each other as shown in Fig.~\ref{fig1}(b). Initially both chains are prepared in their ground states and hence the quantum state of the system is $|GS\ra_L \otimes |GS\ra_R$. A Bell measurement is performed on the first spins of both chains which projects them on one of the following four possible maximally entangled Bell states
\begin{eqnarray} \label{Bell_states}
  |B_0\ra &=& \frac{|\uparrow\downarrow\ra-|\downarrow\uparrow\ra}{\sqrt{2}} \cr
  |B_1\ra &=& \frac{|\uparrow\downarrow\ra+|\downarrow\uparrow\ra}{\sqrt{2}} \cr
  |B_2\ra &=& \frac{|\uparrow\uparrow\ra-|\downarrow\downarrow\ra}{\sqrt{2}} \cr
  |B_3\ra &=& \frac{|\uparrow\uparrow\ra+|\downarrow\downarrow\ra}{\sqrt{2}}
\end{eqnarray}
Since the two chains do not interact any of these four possible outcomes will occur with the probability of 1/4. The symmetry of the system implies that the final entanglement is the same
for all of them and thus we assume that the outcome of the measurement is the singlet $|B_0\ra$. After measurement the first sites of the two chains get entangled and hence at any time $t$ the quantum state of the system can be written as
\begin{equation}\label{psi_ent_t}
    |\psi(t)\ra=2 e^{-iH_Tt} P^{B_0}_{1_L,1_R} |GS\ra_L \otimes |GS\ra_R,
\end{equation}
where $P^{B_0}_{1_L,1_R}=|B_0\ra \la B_0|$ projects the first sites of the two chains (i.e. spins at sites $1_L$ and $1_R$ as depicted in Fig.~\ref{fig2}(b)) into a singlet state $|B_0\ra$, the factor 2 at the beginning of the formula is for normalization and $H_T=H\otimes I+I\otimes H$ is the total Hamiltonian of the system. One can compute the reduced density matrix of the last two sites by tracing out all the rest. The special symmetries of the system and conservation of parity during the evolution implies that
\begin{equation}\label{psi_ent_t}
    \rho_{N_L,N_R}(t)=\frac{1}{2}\left(
                        \begin{array}{cccc}
                          a(t) & 0 & 0 & 0 \\
                          0 & 1-a(t) & b(t)& 0 \\
                          0 & b(t) & 1-a(t) & 0 \\
                          0 & 0 & 0 & a(t) \\
                        \end{array}
                      \right)
\end{equation}
where both $a$ and $b$ are real numbers and can be written as

\begin{widetext}
\begin{eqnarray} \label{a_b_t}
  a(t) = \frac{1}{2} &\{& \la\uparrow\Downarrow| e^{+iHt} |\uparrow_N\ra \la\uparrow_N|  e^{-iHt} |\uparrow\Downarrow\ra \times \la\downarrow\Uparrow| e^{+iHt} |\uparrow_N\ra \la\uparrow_N| e^{-iHt} |\downarrow\Uparrow\ra \cr
       &+& \la\uparrow\Downarrow| e^{+iHt} |\uparrow_N\ra \la\uparrow_N|  e^{-iHt} |\downarrow\Uparrow\ra \times \la\downarrow\Uparrow| e^{+iHt} |\uparrow_N\ra \la\uparrow_N|  e^{-iHt} |\uparrow\Downarrow\ra \cr
       &+& \la\downarrow\Downarrow| e^{+iHt} |\uparrow_N\ra \la\uparrow_N|  e^{-iHt} |\downarrow\Downarrow\ra \times \la\uparrow\Uparrow| e^{+iHt} |\uparrow_N\ra \la\uparrow_N|  e^{-iHt} |\uparrow\Uparrow\ra \cr
       &+& \la\uparrow\Uparrow| e^{+iHt} |\uparrow_N\ra \la\uparrow_N|  e^{-iHt} |\uparrow\Uparrow\ra \times \la\downarrow\Downarrow| e^{+iHt} |\uparrow_N\ra \la\uparrow_N|  e^{-iHt} |\downarrow\Downarrow\ra \cr
       &+& \la\downarrow\Uparrow| e^{+iHt} |\uparrow_N\ra \la\uparrow_N|  e^{-iHt} |\uparrow\Downarrow\ra \times \la\uparrow\Downarrow| e^{+iHt} |\uparrow_N\ra \la\uparrow_N|  e^{-iHt} |\downarrow\Uparrow\ra \cr
       &+& \la\downarrow\Uparrow| e^{+iHt} |\uparrow_N\ra \la\uparrow_N|  e^{-iHt} |\downarrow\Uparrow\ra \times \la\uparrow\Downarrow| e^{+iHt} |\uparrow_N\ra \la\uparrow_N|  e^{-iHt} |\uparrow\Downarrow\ra        \}
\end{eqnarray}
\begin{eqnarray} \label{a_b_t}
  b(t) = \frac{-1}{2} &\{& \la\uparrow\Downarrow| e^{+iHt} |\downarrow_N\ra \la\uparrow_N|  e^{-iHt} |\uparrow\Uparrow\ra \times \la\downarrow\Uparrow| e^{+iHt} |\uparrow_N\ra \la\downarrow_N| e^{-iHt} |\downarrow\Downarrow\ra \cr
       &+& \la\downarrow\Downarrow| e^{+iHt} |\downarrow_N\ra \la\uparrow_N|  e^{-iHt} |\uparrow\Downarrow\ra \times \la\uparrow\Uparrow| e^{+iHt} |\uparrow_N\ra \la\downarrow_N|  e^{-iHt} |\downarrow\Uparrow\ra \cr
       &+& \la\downarrow\Downarrow| e^{+iHt} |\downarrow_N\ra \la\uparrow_N|  e^{-iHt} |\downarrow\Uparrow\ra \times \la\uparrow\Uparrow| e^{+iHt} |\uparrow_N\ra \la\downarrow_N|  e^{-iHt} |\uparrow\Downarrow\ra \cr
       &+& \la\downarrow\Uparrow| e^{+iHt} |\downarrow_N\ra \la\uparrow_N|  e^{-iHt} |\uparrow\Uparrow\ra \times \la\uparrow\Downarrow| e^{+iHt} |\uparrow_N\ra \la\downarrow_N|  e^{-iHt} |\downarrow\Downarrow\ra   \}
\end{eqnarray}
\end{widetext}

\begin{figure} \centering
    \includegraphics[width=9cm,height=4.5cm,angle=0]{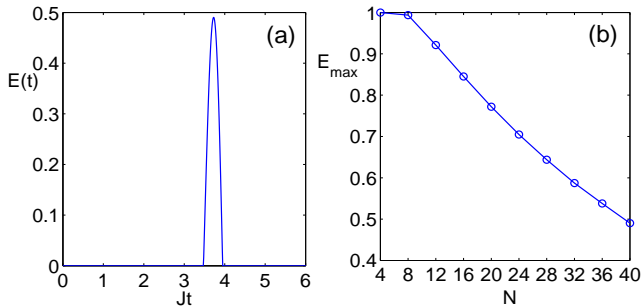}
    \caption{(Color online) (a) Entanglement $E(t)$ as a function of $Jt$ for a chain of length $N=40$ (i.e. $N_L=N_R=20$). (b) The maximal entanglement $E_{max}$ versus length $N$.}
     \label{fig4}
\end{figure}

One can compute the entanglement, quantified by concurrence \cite{concurrence}, between the two qubits which becomes
\begin{equation}\label{concurrence}
    E(t)=max\{0,b(t)-a(t)\}.
\end{equation}
In Fig.~\ref{fig4}(a) the entanglement $E(t)$ is plotted as a function of time. It is worth mentioning that as entanglement propagates in two disconnected chains the distance over which the entanglement is generated at $t=t_{opt}$ is double the distance of state transfer. In Fig.~\ref{fig4}(b) the maximum attainable entanglement $E_{max}=E(t_{opt})$ is plotted versus distance $N$. As it is clear from the figure entanglement decays almost linearly by increasing $N$ with a small slope such that it reaches $E_{max}=0.49$ for a large distance of $N=40$.

\begin{table*} \label{table_1}
\begin{centering}
\begin{tabular}{|c|c|c|c|c|c|c|c|c|c|}
  \hline
  $N (even)$      & 4    & 6    & 8    & 10    & 12    & 14    & 16    & 18    & 20 \\
  \hline
  $F_{av}(even)$ & 0.9991 & 0.9867 & 0.9735 & 0.9604 & 0.9482 & 0.9368 & 0.9264 & 0.9171 & 0.9082 \\
  \hline
  $N (odd)$      & 3 & 5 & 7 & 9 & 11 & 13 & 15 & 17 & 19 \\
  \hline
  $F_{av}(odd)$ & 0.9715 & 0.9526 & 0.9367 & 0.9236 & 0.9118 & 0.9013 & 0.8915 & 0.8834 & 0.8761 \\
  \hline
\end{tabular}
\caption{A comparison between the attainable average fidelity at the optimal time, i.e. $F_{av}(t_{opt})$ between the even and odd chains for the case that the outcome of the measurement is spin up. As the number shows the even chains, with SU(2) symmetry, produce higher fidelity even for slightly longer chains.  }
\par\end{centering}
\end{table*}

\section{Odd Chains} \label{sec6}

So far we have only considered even chains for which the ground state is unique and supports the SU(2) symmetry with total excitation of zero. In contrast, the odd chains have doubly degenerate ground states $|GS_\uparrow\ra$ and $|GS_\downarrow\ra$ that each can be converted to another by applying $\prod_k \sigma_k^x$. In a chain of length $N$, the ground state $|GS_\uparrow\ra$ ($|GS_\downarrow\ra$)  lies in the manifold of parity +1 (-1) in which $(N+1)/2$ number of spins are up (down) and the rest are down (up). In such states there is no SU(2) symmetry and one can split their degeneracy by applying a small magnetic field in the $z$ direction to choose one the ground states. Due to the absence of the SU(2) symmetry the fidelity of state transfer in both restricted and unrestricted basis depends on input parameter $\theta$. So, to quantify the quality of state transfer we consider a system of length $N$ initially prepared in one of its ground states, let say $|GS_\uparrow\ra$. Then a restricted measurement in $\sigma^z$ basis is performed on the first spin of the chain which projects the first qubit on either spin $\uparrow$ or spin $\downarrow$. Depending on the outcome of the measurement a further application of $R_\uparrow$ or $R_\downarrow$ rotates the first spin into $|\psi^{(+1)}\ra$ and initialization process is accomplished. A further time evolution of the system transfers this quantum states through out the chain. Just as before one can trace out the state of all spins but the last one and get the reduced density matrix of the last site $\rho_N(t)$ from which the fidelity is computed just as in Eq.~(\ref{fid_r_ud_t}). To have an input independent quantity one can also average over all possible input states on the surface of the Bloch sphere just as the one in Eq.~(\ref{Fav_ud_t}) to get the average fidelity $F_{av}^{odd}(t)$.

Just as before we consider the first peak of the average fidelity at the optimal time $t_{opt}$. In TABLE II we give a comparison for the average fidelity of even and odd chains versus length $N$ when the outcome of the measurement is spin up. By comparing the values one can realize that the quality of transfer is slightly lower for odd chains. For instant the average fidelity in the odd chain of length $N=19$ is $0.88$ while for a longer even chain of $N=20$ is $0.91$. This means that the SU(2) symmetry of the ground state in the even chains makes the quality of transfer even higher than the slightly shorter chains but with an odd length.

\begin{figure*} \centering
    \includegraphics[width=15cm,height=5cm,angle=0]{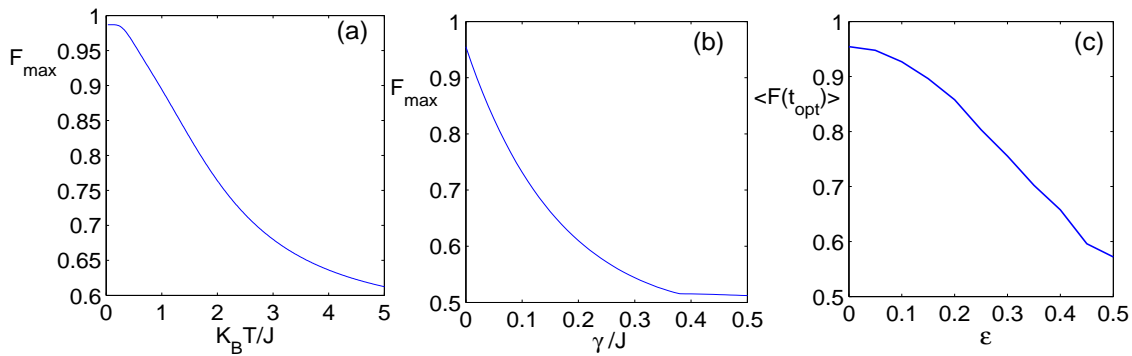}
    \caption{(Color online) The imperfection effects over a chain of length $N=10$: (a) The fidelity $F_{max}$ as a function of dimensionless temperature $K_BT/J$. Thanks to the SU(2) symmetry of the thermal initial state, all projection basis give the same fidelity. (b) The fidelity $F_{max}$ as a function of decoherence coupling $\gamma/J$ when the first qubit is projected into $|+\ra$. (c) The fidelity $\la F(t_{opt})\ra$, averaged over 100 different realizations, in terms of randomness strength $\epsilon$ when the first qubit is projected into $|+\ra$.
     }
     \label{fig5}
\end{figure*}

\section{Imperfections} \label{sec7}

Preparing the system in its anti-ferromagnetic ground state needs cooling to zero temperature which in reality cannot be achieved. Hence, the initial state of the system is inevitably a thermal state of the form $\rho_{th}=\frac{e^{-\beta H}}{Z}$, where $\beta=1/K_{B}T$ in which $T$ is temperature, $K_{B}$ is the Boltzmann constant and $Z$ is the partition function. The transport mechanism is just the same as before. The projective measurement on the first qubit and the following unitary dynamics transfers information across chain just as the case that system has been initialized in its ground state. In fact, the assumption of a unitary evolution is valid only when the thermalization time is much longer than our optimal time $t_{opt}$. In Fig.~\ref{fig5}(a), the maximal attainable fidelity $F_{max}$ is plotted in terms of $K_BT/J$ for a chain of length $N=10$. As SU(2) symmetry remains valid in the thermal initial state, the fidelity is independent of the basis of measurement. As it is evident from the figure, there is a plateau for $F_{max}$ at low temperatures which its width is determined by the finite size energy gap of the system. It is worth mentioning that the optimal time at which the fidelity peaks does not change with temperature which is consistent with the results of \cite{bayat-thermal}.

In practical situations, it is impossible to isolate the system from its environment. To study such effects, we assume that the system is initialized in its ground state and the projective measurement is performed on the first spin just as before. However, we replace the unitary time evolution of the system with a Lindblad type master equation as
\begin{equation} \label{Lindblad}
 \dot{\rho}(t)= -i[H,\rho(t)]+\gamma\sum_{k=1}^{N}\sum_{\mu=1}^2(L^\mu_{k}\rho(t)L^{\mu \dagger}_{k} - \frac{1}{2} \{ L^\mu_{k}L^{\mu \dagger}_{k},\rho(t) \} )
\end{equation}
where $L^1_{k}=\sigma^+_k$ and $L^2_{k}=\sigma^-_k$ are the Lindblad operators which add and subtract spin excitations into the system respectively and the coefficient $\gamma$ represents the coupling with the environment. By tracing out all spins but the last one can compute the fidelity which peaks at $t=t_{opt}$ no matter how strong is the coupling $\gamma$. In Fig.~\ref{fig5}(b) we plot $F_{max}$ as a function of $\gamma$ for chain of length $N=10$ when the first spin is projected to $|+\ra=(|\uparrow\ra+|\downarrow\ra)/\sqrt{2}$. As it is clear, the fidelity goes down by increasing $\gamma$ and stays above $0.75$ even for $\gamma\simeq0.1J$.

Another imperfection is randomness in the coupling of the Hamiltonian as making a uniform chain might be very challenging in some physical realizations. This means that in the Hamiltonian of Eq.~(\ref{H}) we have $J_{k}=J(1+\delta_{k})$, where, $\delta_{k}$ is a dimensionless random number with a uniform distribution in the interval $[-\epsilon,\epsilon]$. In fact,  $\epsilon$ determines the strength of randomness in the couplings. We fix the optimal time to be $t_{opt}$, determined from the uniform chain (i.e.  $\epsilon=0$), as the real time at which fidelity peaks depends on all couplings $J_k$'s. We then average the fidelity $F(t_{opt})$ over several different realizations (we did for 100) of the system for a fixed $\epsilon$. In Fig.~\ref{fig5}(c) we depict the fidelity $\la F(t_{opt}) \ra$ averaged over 100 different realization as a function of $\epsilon$ when the first qubit is projected into the state $|+\ra$. It is seen that although the average fidelity decreases by increasing the randomness the mechanism shows a relatively high resistance against this destructive effect as fidelity remains above $0.85$ even for twenty percent of randomness (i.e. $\epsilon=0.2$).

\begin{figure} \centering
    \includegraphics[width=8cm,height=6cm,angle=0]{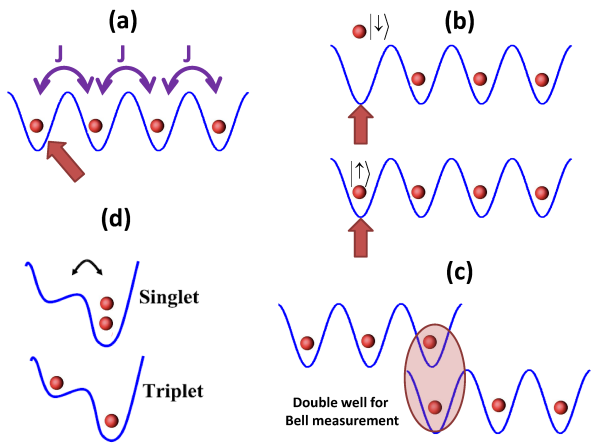}
    \caption{(Color online) (a) Cold atoms in an optical lattice prepared in a Mott insulator phase with exactly one atom per site realizes the Heisenberg Hamiltonian of Eq.~(\ref{H}). A local focused laser beam is used to manipulate the first qubit for both the gate operation and spin measurement. (b) The single qubit measurement is accomplished by a perpendicular focused laser beam which applies a strong radiation pressure to the state $|\downarrow\ra$ and leaves the atom unaffected if its quantum state is $|\uparrow\ra$. (c) Two parallel arrays in a two dimensional optical lattice are used for entanglement distribution in which a Bell measurement is needed. (d) Bell measurement is fulfilled by tilting the optical lattice such that the singlets tend to occupy a single site and triple pairs remain separated.}
     \label{fig6}
\end{figure}

\section{Application for Optical Lattices} \label{sec8}

The proposed mechanism is most suitable for realization in optical lattices in which an array of cold atoms in their Mott insulator phase sit in the minimums of a periodic potential, formed by counter propagating laser beams, as shown in Fig.~\ref{fig6}(a). In the limit of high on-site energy the double occupancy is prohibited and the interaction between atoms is effectively explained by a spin Hamiltonian \cite{Lukin}. Changing the intensity of the laser beams tunes the tunneling rate of the atoms and thus controls the exchange coupling of the spin chain globally. In two or three dimensional lattices by tuning the intensity of the corresponding laser beams one can independently control the coupling of the atoms in each dimension globally. Recently, local addressability of the atoms have also been possible in optical lattices \cite{Bloch-single-site1,Bloch-single-site2}, makes local measurements and spin rotations, the two essential ingredients of our proposal accessible. Using such local operations the propagation of a single \cite{Bloch-spin-wave} and double  \cite{Bloch-magnon} spin flips in a ferromagnetic chain have been experimentally observed.

To perform spin measurement on a single site one can use the techniques developed in Ref.~\cite{Meschede}. In that methodology an intense perpendicular leaser beam is focused to the target atom and couples one of the atomic levels which represents $|\downarrow\ra$ to one of the excited states. This generates a strong radiation pressure which pushes the atom out of the lattice only when atom is in state $|\downarrow\ra$ and does not affect it otherwise. This leaves the site empty if its atom is in state $|\downarrow\ra$ and full if the atom is in state $|\uparrow\ra$ as it is shown schematically in Fig.~\ref{fig6}(b). So, the result of the measurement is revealed through a following fluorescent picture to see whether the atom is still sitting in its initial position (projecting to $|\uparrow\ra$) or has gone (projecting to $|\downarrow\ra$). Notice that in this technique by probability of 1/2, for which the atom is in the state $|\downarrow\ra$ and thus leaves the lattice, the protocol fails which reduces the rate of communication by half. This means that if a two dimensional optical lattice is used to provide several equivalent parallel noninteracting spin chains (just as the one for ferromagnetic case in Refs. \cite{Bloch-spin-wave,Bloch-magnon}) and the measurement is performed instantaneously on all the first qubits of parallel chains only half of them can be used to extract final information as there will be no hole in those chains and the rest should be discarded.

Apart from single qubit measurement we also need to perform unitary operations (such as $R_\uparrow$ and $R_\downarrow$ in Eq.~(\ref{R1_R2})) to accomplish the initialization and encoding information. To apply such unitary operators on the target atom (i.e. site $1$) a focused laser beam is exploited to generate Rabi oscillation between the qubit levels as shown in Fig.~\ref{fig6}(a). This local operation is much quicker ($\sim 10\mu s$) \cite{Meschede} than the time evolution of the system ($\sim 1-10$ ms) \cite{Bloch-spin-wave,Bloch-magnon} and can be considered as a sudden action.
To have a pure local gate operation and avoid affecting the neighboring qubits one may apply a weak magnetic field gradient \cite{Meschede}, which splits the hyperfine levels of all qubits position dependently, or use a tightly focused laser beam \cite{Bloch-single-site2} to only split the hyperfine levels of the target atom. So then a microwave pulse, tuned only for the target qubit, operates the gate locally as it has been realized in Refs.~\cite{Bloch-single-site2,Meschede}.
For instance, a weak magnetic field gradient of $27.4 Gcm^{-1}$ is enough for applying  $\sigma^x$ on a target qubit with a pulse of duration $10\mu s$ without affecting the neighboring sites~\cite{Meschede}.

According to the proposed mechanism for entanglement distribution a Bell measurement on the first qubits of the two chains is essential for initializing the system. We consider a geometry, shown in Fig.~\ref{fig6}(c), in which two arrays of atoms sit in two parallel rows with the first atoms recite in the neighboring sites. To perform the Bell measurement we first raise the barriers between the atoms to switch off the interactions along the chains (i.e. $J=0$ in both spin chains). We use the fact that the energy levels for the singlet and triple pairs are different in a single well such that the singlet state is lower in energy. To operate the Bell measurement one has to tilt the lattice adiabatically such that the atoms in the right chain tunnel into the next row and sit along the left chain. Though, the atom in the first site of the right chain has to compensate an extra on site energy $U$ for its tunneling as its target site is already occupied by the first atom of the left chain. If the amount of tilting is tuned to be resonant \emph{only} with the singlet state of two atoms in the doubly occupied site then the double occupancy occurs only for the singlet state as shown in Fig.~\ref{fig6}(d). As the other Bell states are off resonant and energetically inaccessible, the double occupancy never occurs for such states. A further florescent picture of the system, which can be done without disturbing the internal states \cite{Gibbons-Nondestructive}, will determine the number of atoms in the first site and reveals if the two atoms are in a singlet state or not. A backward adiabatic evolution (i.e. returning the lattice back to its normal) restores all the atoms into their initial position while the first spins are either projected to singlet $|B_0\ra$ or one of the three other Bell states. If the output of the projecting measurement is singlet $|B_0\ra$ (its probability is 1/4) then initialization is complete and by decreasing the horizontal barriers along the chains the propagation begins. On the other hand if the result is not $|B_0\ra$ then the density matrix of the two qubits is an equal mixture of all other Bell states (its probability is 3/4). One then can apply $\sigma_z$ to the atoms in site $1_L$ (or $1_R$) in order to convert the $|B_1\ra$ part of the mixture into $|B_0\ra$ and repeat the adiabatic tilting to see if the projection to singlet is accomplished or not. This time the probability of success increases to 1/3. In the case of failure the state of the two atoms become a mixture of $|B_2\ra$ and $|B_3\ra$ which a local unitary operation $\sigma_y$ transforms these two states into $|B_0\ra$ and $|B_1\ra$ respectively. An extra repeating of the adiabatic tilting either directly gives a singlet state $|B_0\ra$ (with the probability of 1/2) for the pair or project them into $|B_1\ra$ (again with probability of 1/2) which then can be transformed to $|B_0\ra$ locally. Hence, at the worst case the adiabatic tilting of the lattice has to be done three times for the initialization. Then by letting the system to evolve one can generate entanglement between the distant atoms at both sides of the system.

\section{Conclusion} \label{sec9}
In this paper we put forward a timely proposal for quantum communication in anti-ferromagnetic Heisenberg Hamiltonian using only local operations for encoding the information. This harnesses the \emph{intrinsic entanglement} of the system for inducing dynamics via a single site quantum measurement. As the outcome of measurement is ultimately random a following unitary operation which is determined by the outcome of the measurement is essential for encoding the information within the intrinsically entangled ground state of the system. By finishing the encoding procedure system is left to evolve freely and after a certain time (set by the length $N$ and the strength of the exchange coupling $J$) information reaches the receiver site which can be taken for further computational process. The quality of state transfer remains above the threshold limit for chains up to length $N\sim 50$ while system is not engineered and no extra modulation is needed.
In comparison with the widely studied attaching scenarios, our proposed mechanism not only introduces a different encoding of quantum states into a many-body system which harnesses the intrinsic entanglement of the ground state but also provides a new way for inducing quantum quench in such systems. From the perspective of quantum communication our measurement induced transport gives higher average fidelity and does not need local control over interaction at least on the sender site. One application of our proposal can be information router in which the quantum state is prepared at a particular site to simplify the fabrication and then is distributed among multiple users. In fact, an immediate generalization of our idea is to design an information router based on the proposed measurement induced transport which has to be pursued in a separate project.  Alternatively, one may see our protocol as remote quantum state preparation \cite{Bennet-remote} in which a known quantum state is generated remotely at the output via the free evolution of a many-body strongly correlated system. In addition, we considered several imperfections which may arise in different realizations including thermal fluctuations, interaction with environment and the effect of random couplings.

Since the encoding of information and performing the quantum quench in the system is done by only local operations the proposed mechanism is most suitable to be realized in optical lattices. The recent experiments for spin wave propagation \cite{Bloch-spin-wave} and transferring magnon bound states \cite{Bloch-magnon} show that all the ingredients we need is already available in the laboratory. Based on these new achievements, our proposal is just timely for being pursued in experiments and indeed can be realized with current technology.

{\em Acknowledgements:-} Discussions with Sougato Bose, Leonardo Banchi and Bedoor Alkurtas are warmly acknowledged. AB thanks EPSRC grant $EP/K004077/1$.

\appendix

\section{SU(2) symmetry of the Heisenberg Hamiltonian}

The SU(2) group and its corresponding SU(2) Lie algebra are fully determined by the Pauli operators as the generators of the algebra.
Any element of the SU(2) group in its $2\times 2$ representation can be written as
\begin{equation}\label{SU2_U}
    \mathcal{U}(\alpha,\widehat{n})=e^{i \alpha \overrightarrow{\sigma}.\widehat{n}}
\end{equation}
where $\alpha$ is a real number and $\widehat{n}$ is a unit vector in the three dimensional space. The SU(2) symmetry of the Heisenberg Hamiltonian of Eq.~(\ref{H}) means
\begin{equation}\label{SU2_symmetry}
    \mathcal{U}(\alpha,\widehat{n})^{\otimes N} H \mathcal{U}^\dagger(\alpha,\widehat{n})^{\otimes N}=H.
\end{equation}
Being a spin singlet, the ground state of the Hamiltonian for even $N$ is also invariant, up to an irrelevant global phase, under the action of $\mathcal{U}(\alpha,\widehat{n})^{\otimes N}$ such that
\begin{equation}\label{SU2_GS}
    \mathcal{U}(\alpha,\widehat{n})^{\otimes N}|GS\ra=e^{i\beta}|GS\ra
\end{equation}
where $\beta$ is a global phase. One can show that the $R_\uparrow$ operator, defined in Eq.~(\ref{R1_R2}), is an element of SU(2) group as $R_\uparrow=\mathcal{U}(\alpha^*,\widehat{n}^*)$ for a particular choice of
\begin{eqnarray}\label{SU2_theta-n}
    \alpha^*= \theta/2 \h
    \widehat{n}^*= \left (\sin(\phi), \cos(\phi), 0 \right )
\end{eqnarray}
where $\theta$ and $\phi$ are the qubit parameters in Eq.~(\ref{psi_p1}). Following the Eqs.~(\ref{SU2_symmetry}) and (\ref{SU2_GS}), this implies that
 \begin{equation}\label{SU2_R_GS_H}
    R_\uparrow^{\dagger \otimes N} |GS\ra=e^{i\beta}|GS\ra, \h R_\uparrow^{\dagger \otimes N} H R_\uparrow^{\otimes N}=H.
\end{equation}

We now have all the ingredients to prove that the fidelity of the unrestricted basis strategy is independent of the qubit parameters $\theta$ and $\phi$. Let's assume that the projection is made in the basis of $\{ |\uparrow\ra,|\downarrow\ra \}$ (which corresponds to $\theta=0$ while $\phi$ is arbitrary) and the outcome of the measurement is $|\uparrow\ra$ (namely the $+1$ solution). We show that the fidelity is in fact the same for all other values of $\theta$ and $\phi$ provided that the measurement outcome is $+1$. The time evolution of the system can be written as
\begin{equation}\label{SU2_psi_t}
    |\Psi(t)\ra=\sqrt{2} e^{-iHt} P_1^{\uparrow} |GS\ra
\end{equation}
where $P_1^{\uparrow}=|\uparrow_1\ra \la \uparrow_1|$ is the projection on the first qubit. The fidelity then can be written as
\begin{eqnarray}\label{SU2_Fp} \nonumber
    F^+_u (\theta=0, \phi)&=& \la \Psi(t)| P_N^\uparrow |\Psi(t) \ra  \cr
    &=& 2 \la GS| P_1^\uparrow e^{+iHt} P_N^\uparrow e^{-iHt} P_1^{\uparrow} |GS \ra .
\end{eqnarray}
Using the equalities in Eq.~(\ref{SU2_R_GS_H}) one can insert $R_\uparrow^{\otimes N}$ and its Hermitian conjugate on both sides of the time evolution operators and apply it to the ground state $|GS\ra$ without changing the fidelity. With a straight forward calculation one gets
\begin{eqnarray}\label{SU2_Fp2}
    F^+_u (\theta=0, \phi)&=&  2 \la GS| P_1^{(+1)}  e^{+iHt} P_N^{(+1)} e^{-iHt} P_1^{(+1)} |GS \ra \cr
    &=& F^+_u (\theta, \phi).
\end{eqnarray}
where we have used the fact that
\begin{equation}\label{SU2_Proj_p_up}
    P_k^{(+1)}=R_\uparrow P_k^\uparrow R_\uparrow^\dagger, \h k=1,N.
\end{equation}
By arriving to the Eq.~(\ref{SU2_Fp2}) the proof is complete.

\end{document}